# Variable-Temperature Inelastic Light Scattering Spectroscopy of Magnons and Phonons in Nickel Oxide

M.M. Lacerda[1,2], F. Kargar[1,3], E. Aytan[1,3], R. Samnakay[1,3], B. Debnath[4], J. X. Li[5], A. Khitun[3,6], R. K. Lake[4,6], J. Shi[5,6], A. A. Balandin[1,6,*]

[1]Nano-Device Laboratory (NDL) and Phonon Optimized Engineered Materials (POEM) Center, Department of Electrical and Computer Engineering, University of California – Riverside, Riverside, California 92521 USA

[2]Campus Duque de Caxias, Universidade Federal do Rio de Janeiro, RJ, Brasil

[3]Materials Science and Engineering Program, Bourns College of Engineering, University of California – Riverside, Riverside, California 92521 USA

[4]Laboratory for Terascale and Terahertz Electronics (LATTE), Department of Electrical and Computer Engineering, University of California – Riverside, Riverside, California 92521 USA

[5]Department of Physics and Astronomy, University of California – Riverside, Riverside, California 92521 USA

[6]Spins and Heat in Nanoscale Electronic Systems (SHINES) Center, University of California – Riverside, Riverside, California 92521 USA

[*]Corresponding author (AAB): balandin@ee.ucr.edu

# Abstract

We report results of an investigation of the temperature dependence of the magnon and phonon frequencies in NiO. A combination of Brillouin-Mandelstam and Raman spectroscopies allowed us to elucidate the evolution of the phonon and magnon spectral signatures from the Brillouin zone center (GHz range) to the second-order peaks from the zone boundary (THz range). The temperature-dependent behavior of the magnon and phonon bands in the NiO spectrum indicates the presence of antiferromagnetic (AF) order fluctuation or a persistent AF state at temperatures above the Néel temperature ($T_N$=523 K). Tuning the intensity of the excitation laser provides a method for disentangling the features of magnons from acoustic phonons without the application of a magnetic field. Our results are useful for interpretation of the inelastic-light scattering spectrum of NiO, and add to the knowledge of its magnon properties important for THz spintronic devices.

Nickel Oxide (NiO) has been extensively studied in the past for a variety of applications in catalysts, gas sensors, electrochemical films, battery electrodes and photo-devices[1–6]. Recently, NiO has attracted a renewed, and fast growing, interest as an antiferromagnetic (AF) electrically insulating material for applications in spintronic devices, particularly those operating at THz frequencies.[7] NiO has also been used for spin current enhancement as an intermediate nanometer-scale layer in magnetic material – NiO – heavy metal heterostructures[8,9]. In general, utilization of spin currents instead of electric currents has the advantage of reducing Joule heating. The latter constitutes a promising approach for the next generation of devices for low-power-dissipation information processing[10,11]. Further development of spintronics requires better understanding of the physics of magnons and phonons and their interactions in AF materials.

Despite a number of prior studies of the Raman spectrum of NiO[12–17] there is substantial discrepancy among the reported values for the phonon and magnon frequencies, and between theory and experiment[18–21]. Usually, it is difficult to distinguish magnon peaks from phonon peaks, particularly when they appear as a broad band, which is a combinations of both[11,22]. Application of a magnetic field – a conventional approach for identification of magnons – is difficult for NiO due to the large field strength required for inducing measurable magnon peak shifts. It has been reported[23] that application of a magnetic field with the intensity $B$=7 T to NiO resulted in the Raman magnon frequency change of less than ~3 cm$^{-1}$. The magnetic ordering affects phonon energies via spin – phonon and spin – lattice coupling[24,25]. Calculations of the phonon band structure, which do not include effects of spin ordering, predict phonon energies that differ from the experiments[19,26]. Similarly, magnon dispersion, calculated from the Heisenberg model for spins located on the same and adjacent crystalline planes without inclusion of the spin-lattice interactions, leads to discrepancy with experimental data. For example, the most studied two-magnon (2M) peak in the Raman spectrum of NiO, which is observed at ~1500 cm$^{-1}$ at room temperature (RT), has been theoretically predicted[17,20] in the range of 1800 cm$^{-1}$ – 1900 cm$^{-1}$. The difficulties in interpretation of the inelastic-light scattering spectra of NiO are not limited to the optical phonon frequencies (~200 cm$^{-1}$ – 2000 cm$^{-1}$) measured with Raman spectroscopy. They are also present in the acoustic phonon frequency range (2 GHz – 400 GHz; 1 THz = 33.37 cm$^{-1}$) measured with Brillouin – Mandelstam spectroscopy (BMS)[23,27]. The low signal-to-noise ratios for BMS peaks of magnons from the Brillouin zone (BZ) center coupled with the overlap in their frequency with phonons, and discrepancy with the theoretical predictions further complicate the peak assignment.

In this Letter, we report results of a combined Raman and BMS study of NiO crystals over a wide temperature range. Raman spectra were measured using the cold-hot cell with the temperature changing from RT to 873 K. The temperature rise in BMS experiments was achieved by increasing the excitation laser intensity to 250 mW on the sample surface, which resulted in local temperatures exceeding 700 K. Based on the results of BMS studies, we propose a simple method for distinguishing BZ-center (Γ point) magnon peaks from acoustic phonon peaks using the excitation-laser power dependence of the magnon peak intensity. We have also established that the intensity of the BZ-edge two-magnon peak in the Raman spectrum of NiO is strongly suppressed above the Néel temperature where the crystal undergoes reconstruction from the AF to the paramagnetic state[12,22]. However, the 2M peak does not disappear completely at T~600 K (above $T_N$=523 K) indicating AF order fluctuations or persistent AF ordering. Heating NiO crystals above 870 K results in irreversible changes in the Raman spectrum related to structural modification of the samples.

The single crystal nature of bulk NiO [111] samples selected for this study has been confirmed by X-ray diffraction (XRD) measurements. NiO has a an A-type AF crystalline structure consisting of ferromagnetically (FM) aligned (111) planes that are AF aligned in respect to each other[18,28,29]. Below the Néel temperature, the AF ordering is accompanied by a slight rhombohedral distortion[29].

The Raman measurements were performed in the backscattering configuration under laser excitation of λ = 488 nm (2.54 eV). All measurements were performed with the samples placed inside a cold-hot cell under argon atmosphere to ensure that no oxidation occurred at elevated temperatures. The excitation power was kept at 2 mW to avoid local heating. Under such conditions the temperature of the sample is the same as set by the cold-hot cell. Raman spectra were accumulated during both the heating and cooling cycles. For accurate measurement of the characteristic magnon and phonon frequencies all spectra were fitted with Voigt functions.

Figures 1 (a) and (b) illustrate the evolution of the Raman spectrum of (111) NiO during the heating and cooling cycles, respectively. The fact that the spectrum changes are completely reversible when the samples are heated up to ~600 K and cooled back to RT proves that the Raman spectrum changes are due to intrinsic phonon and magnon properties of NiO, and they are not related to any surface oxidation or contamination. The Raman bands measured at RT (upper panel in Figure 1 (a)) are in agreement with previous studies of bulk NiO single crystals[13]. One can recognize the transverse optical (TO) phonon mode at ~403 cm$^{-1}$ and the longitudinal optical (LO) phonon mode at ~ 520 cm$^{-1}$. The two-phonon excitations observed at ~720 cm$^{-1}$, 899 cm$^{-1}$ and 1100 cm$^{-1}$ correspond to 2TO, TO + LO and 2LO modes, respectively. The pronounced two-magnon peak has the frequency of 1500 cm$^{-1}$. The 2M peak originates from two counter-propagating magnons with the wave vectors from the opposite edges of BZ allowed by the momentum conservation[30]. The 2M band is Raman active due to an induced non-zero net dipole momentum at X and Z symmetry points in the BZ[20]. Its energy approximately corresponds to twice the magnon energy at the BZ boundary[18,20].

[Figure 1]

One can see substantial changes in Raman spectrum of NiO as the temperature increases and crosses the Néel temperature. Near $T_N$ = 523 K, the characteristic 2M band broadens, shifts to lower frequencies and decreases in intensity. The intriguing observation is that 2M signatures are persistent at temperatures well above $T_N$.[31] Similar features have been observed in other AF materials such as RbMnF$_3$[32], MnF$_2$ and KNiF$_3$[33]. The nature of the AF phase transition in NiO, *i.e.* first-order *vs*. second-order, has been debated[34–36]. Several theoretical studies suggested the first-order transition[35,36]. More recent high-temperature neutron diffraction measurements demonstrated persistent AF spin correlations up to 800 K suggesting that the AF phase transition in NiO is indeed continuous[34]. Our Raman data with clear signatures of the magnon band above the Néel temperature constitute an independent proof of this fact using a different experimental technique. Another observation from Figure 1 is that the temperature dependence of the 2LO phonon band is much weaker that than of 2M band. The 2LO phonon frequency remains near 1100 cm$^{-1}$ in the examined temperature range.

As the next step in our study, we increased the temperature substantially above $T_N$ to T=873 K. At temperatures above ~800 K, the Raman spectrum of NiO loses reversibility (see Figure 2 (a)). At T=617 K and T=673 K the LO, LO+TO and 2LO phonon modes are still measurable and their spectral positions are consistent with the data presented in Figure 1. However, in addition to these modes, two new peaks at ~1345 cm$^{-1}$ and 1580 cm$^{-1}$ emerge. The first-order phonon modes are not expected in the paramagnetic state of NiO (above $T_N$) unless there are parity-breaking imperfections in the sample[12] or a persistent AF state above $T_N$. The presence of the 2LO band and traces of TO and LO peaks above $T_N$ is another indicator of residual AF ordering, in line with our observations for the 2M band. From T=673 K up to T=873 K, we observe decreasing intensity of the 2LO phonon band and increasing intensity of new peaks at 1345 cm$^{-1}$ and 1580 cm$^{-1}$. The Raman spectrum changes above T~800 K are irreversible, i.e. the RT spectrum is not restored upon cooling down the sample. XRD analysis confirms that NiO sample undergoes structural changes and becomes polycrystalline (see Figure 2 (b)). The transition to polycrystalline NiO can explain the appearance of new peaks in the Raman spectrum measured significantly above $T_N$. However, their exact assignment is not possible at the moment and requires additional investigation.

[Figure 2]

We now analyze the temperature effects on magnons and acoustic phonons near the Γ point in the BZ. BMS experiments were carried out in backscattering geometry using a solid-state diode pumped laser operating at $\lambda$=532 nm. The laser light was focused on the samples through a lens with $NA$=1.4. The scattered light was collected with the same lens and directed to the high-resolution six-pass tandem Fabry-Perot interferometer. During the experiment, the laser light was focused on the sample at fixed incident angle of 20° with respect to the normal to the sample's surface. The probing phonon and magnon wave-vector in this experiment is $q_B = 4\pi n/\lambda$, where $n$ is the refractive index of NiO at the laser excitation wavelength ($n$~2.4-2.5 [Ref. 37]). Assuming $n$~2.4, the probing wave-vector for bulk phonons and magnons is 56.7 μm$^{-1}$. The laser power on the sample surface was varied from ~60 mW up to ~260 mW. Changing the laser power one can effectively control the local temperature of the sample.

In Figure 3 (a) we present BMS spectrum of NiO with two prominent peaks at 37.9 GHz (1.26 cm$^{-1}$) and 65.3 GHz (2.18 cm$^{-1}$), which correspond to the transverse acoustic (TA) and longitudinal acoustic (LA) phonon branches, respectively. A broad shoulder close to the TA phonon frequency is considered to be a "zero-frequency" magnon band in some studies [27,38]. However, clear experimental evidence for the nature of this band is still missing. Figure 3 (b) shows the BMS spectra over a larger free spectrum range under different incident laser power. At the laser power of 58 mW (black curve), there is a broad peak at ~356 GHz (11.88 cm$^{-1}$), which has been assigned as the zone-center magnon[38]. This is the only magnon band, which was clearly observed in our BMS experiments. This band has been reported in other studies as well, although there is an uncertainty in its frequency[38,39]. It should be noted that the number and spectral positions of the zone-center magnons in NiO have been subjects of debates[15,17,18,20,23,27,38]. The most prominent zone-center magnon is usually observed at ~1 THz[15,17,27,39]. The changes in the magnon peak position with temperature further complicates the peak assignment.

[Figure 3]

Using the peak at ~356 GHz as an example, we now demonstrate a convenient method for distinguishing magnon signatures, which can be readily used in BMS experiments even without cold-hot cells for external temperature control. Figure 3 (b) shows the evolution of the peak with the laser power, $P$, increasing from 58 mW to 250 mW, which results in the corresponding increase of the local temperature. The intensity of the ~356 GHz peak has been normalized to the intensity of the LA phonon peak at ~65 GHz. The position of the LA phonon does not change with increasing power, which indicates that the change in the refractive index, $n$, in this temperature range is negligible (the probing wave-vector in a BMS experiment is a linear function of $n$). At the same time, with increasing laser power, the frequency and the intensity of the ~356 GHz peak decreases and finally, at $P$=250 mW, the peak disappears completely. The frequency of this peak versus $P$ is plotted in Figure 3 (c). We argue that the peak disappearance confirms its magnon origin, and it is due to the local temperature increase above $T_N$ as the laser power on the sample reaches high values. The LA phonon peak hardly changes its frequency and intensity with $P$, and it remains prominent at this high excitation power level (see inset to Figure 3 (c)). Heating the sample with the excitation laser above $T_N$ provides a convenient method for assignment the magnon peaks in NiO and other AF materials without the use of magnetic field and cold-hot cell.

To verify the fact that temperature above $T_N$ is achieved in this experiment, we simulated heat diffusion in a NiO crystal using the finite-element method implemented with COMSOL. Figure 4 shows the maximum temperature of the hotspot and the minimum temperature inside the sample at different incident laser powers. One can see that as the applied laser power increases to ~130 mW, the hotspot temperature reaches the Néel temperature, although the overall sample temperature remains below $T_N$. At ~250 mW, the temperature of the whole sample increases beyond the Néel temperature. The temperature distribution profile in the sample is shown in the inset to Figure 4. The distribution near the hot spot area (red color) depends on the Gaussian profile of the laser power and exponential decay of the light intensity inside NiO (z axis). The sample temperature reaches steady-state within ~1 minute, which is much smaller than actual heating time in the experiment.

[Figure 4]

In summary, the zone-boundary magnon bands and zone-center TO, LO phonons have been analyzed over a wide temperature range extending beyond 870 K. The Raman and BMS spectra indicate that spin correlations resulting in persistent AF ordering or AF order fluctuations exist well above the Néel temperature. An elegant method for disentangling the features of magnons from acoustic phonons in the BMS spectrum without the application of a magnetic field is also demonstrated.

*Acknowledgements*

The work at UC Riverside was supported as part of the Spins and Heat in Nanoscale Electronic Systems (SHINES), an Energy Frontier Research Center funded by the U.S. Department of Energy, Office of Science, Basic Energy Sciences (BES) under Award # SC0012670. M.M.L.

also acknowledges Conselho Nacional de Desenvolvimento a Pesquisa (CNPq) and the program Ciencias sem Fronteiras for financial support during her research at UC Riverside.

# FIGURE CAPTIONS

**Figure 1:** (a) Evolution of Raman spectrum of NiO at different temperatures during the heating cycle. (b) Raman spectra of NiO during the cooling cycle. The changes in Raman spectrum are fully reversible within this temperature range. Note traces of two-magnon peak (2M) at T=577, which is above Néel temperature $T_N$=523 K.

**Figure 2:** (a) Raman spectrum of NiO over an extended temperature range reaching T=873 K. The changes in Raman spectrum of NiO heated above 600 K are irreversible. Note appearance of new peaks at T=617 K. (b) XRD spectrum of NiO before (black curve) and after (red curve) heating to T=873 K. The changes in XRD spectrum are in line with Raman data and confirm structural changes in NiO.

**Figure 3:** (a) BMS spectrum of NiO with pronounced TA and LA phonon peaks. (b) BMS spectrum of NiO at different excitation laser power on a sample surface. The one-magnon peak at ~356 GHz shifts to lower frequency and disappears completely as the temperature increases above $T_N$ owing to the laser heating. (c) The change in the magnon peak position with the laser power. The inset shows LA phonon peak as a function of laser power. The results demonstrate that the magnon signatures can be conveniently identified from their laser power dependence without the use of magnetic field or cold-hot cells.

**Figure 4:** Calculated temperature as a function of the excitation laser power for a given NiO sample. The Néel temperature TN is shown by horizontal line. The red curve corresponds to the highest temperature in the hot spot while the blue curves provides the lowest temperature on the back of the sample. Inset presents temperature distribution in the simulated NiO sample. The laser power is set as100 mW. The calculations support the conclusion that $T>T_N$ are achieved in the BMS experiments with variable laser power.

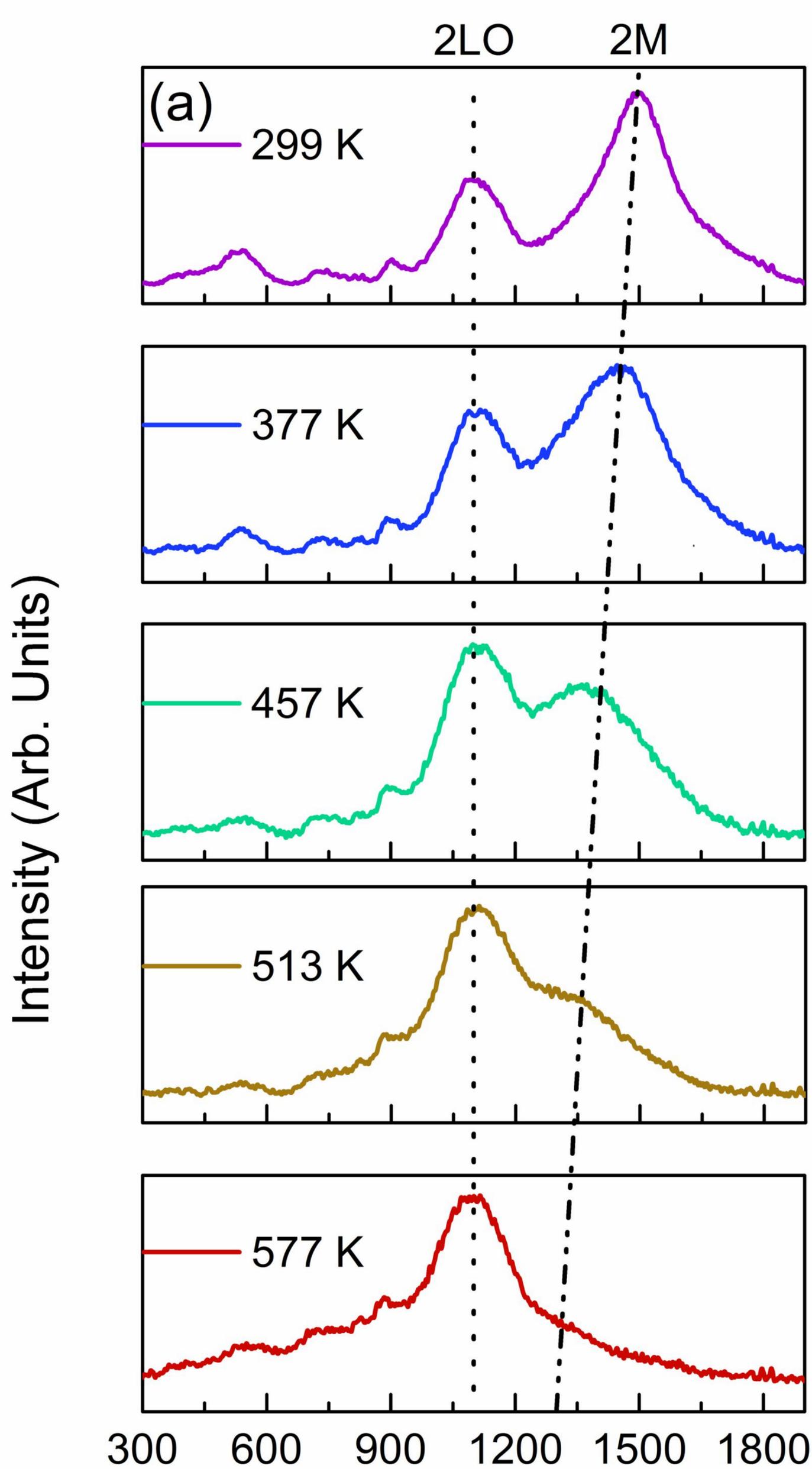
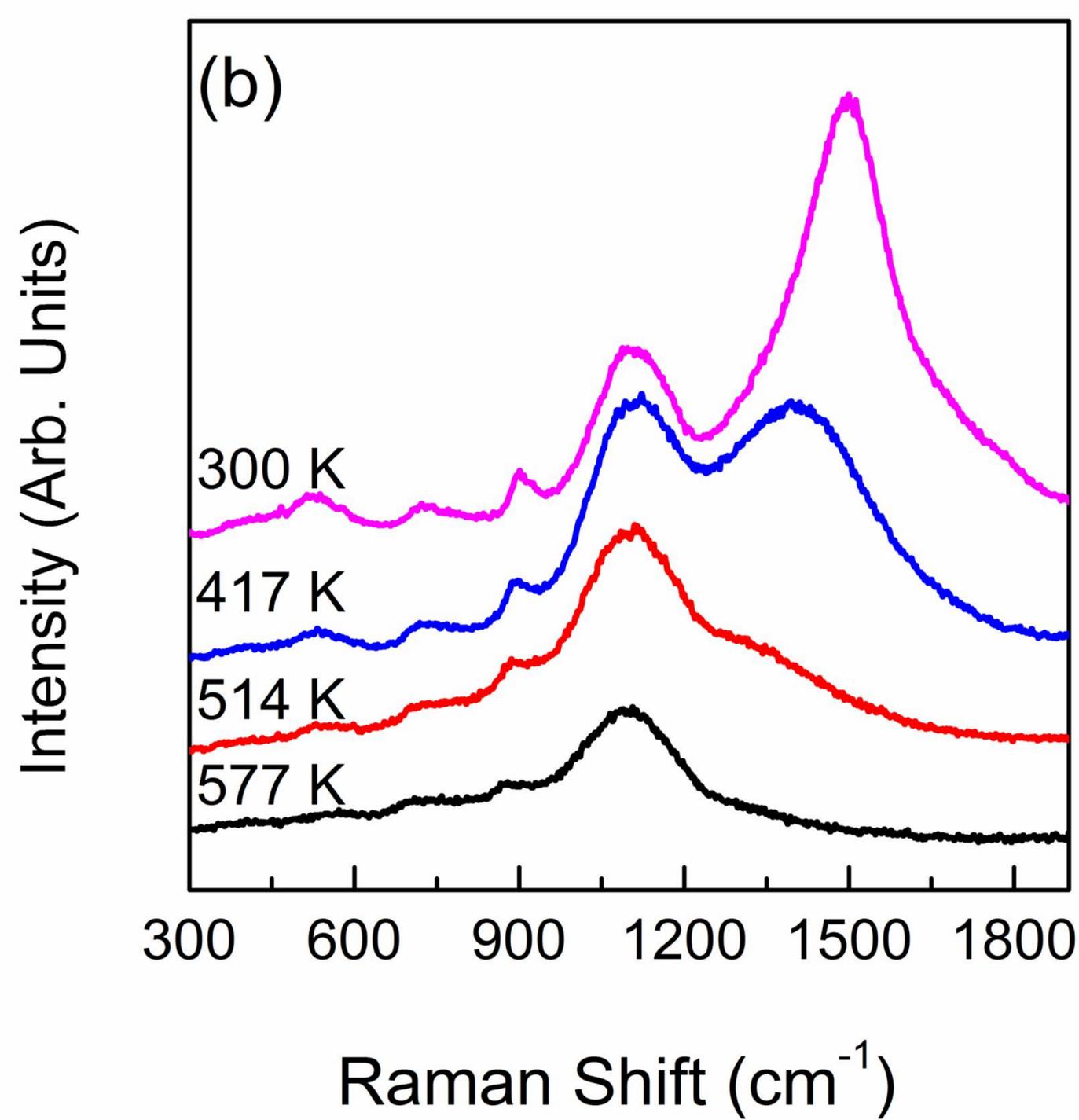

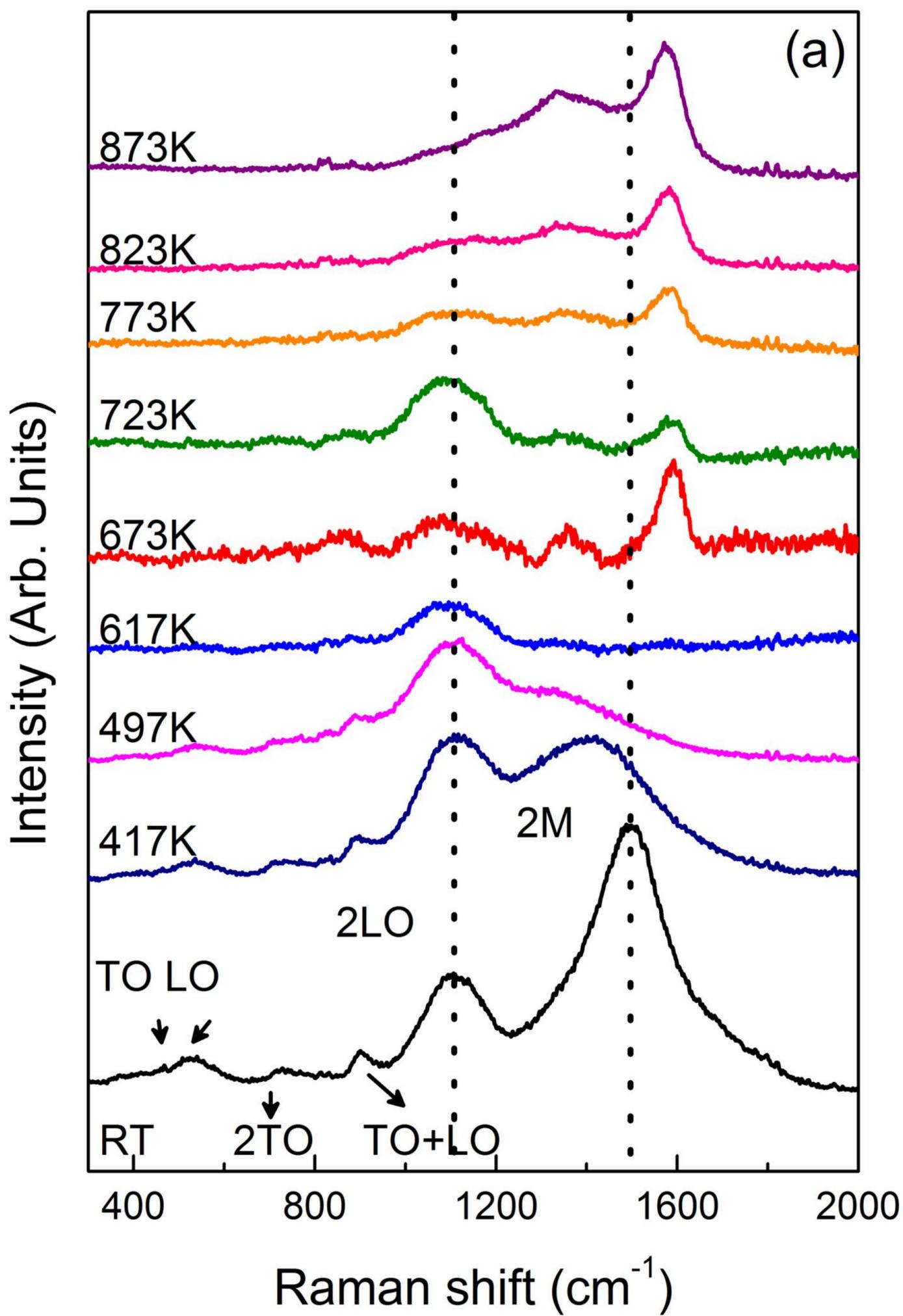

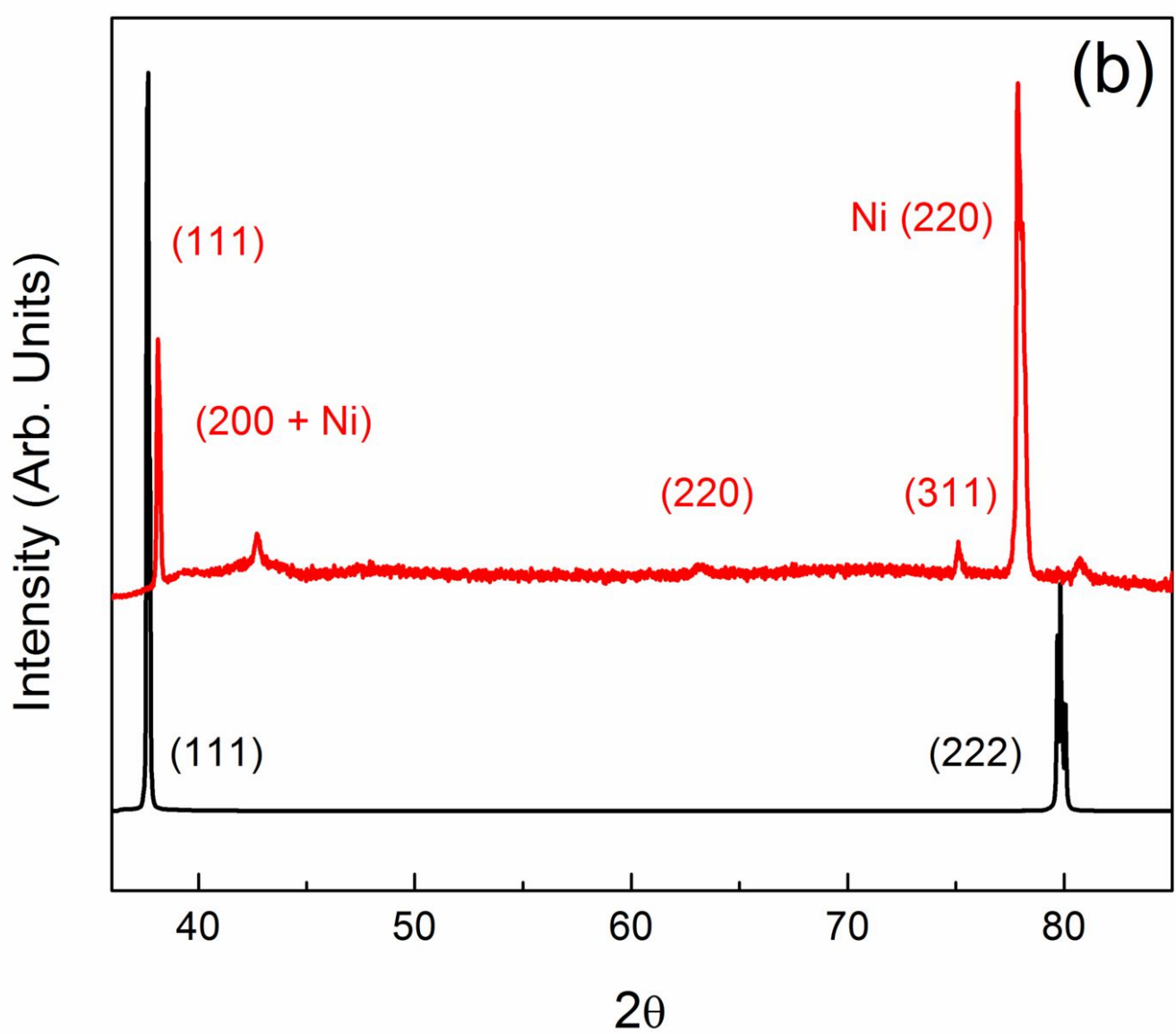

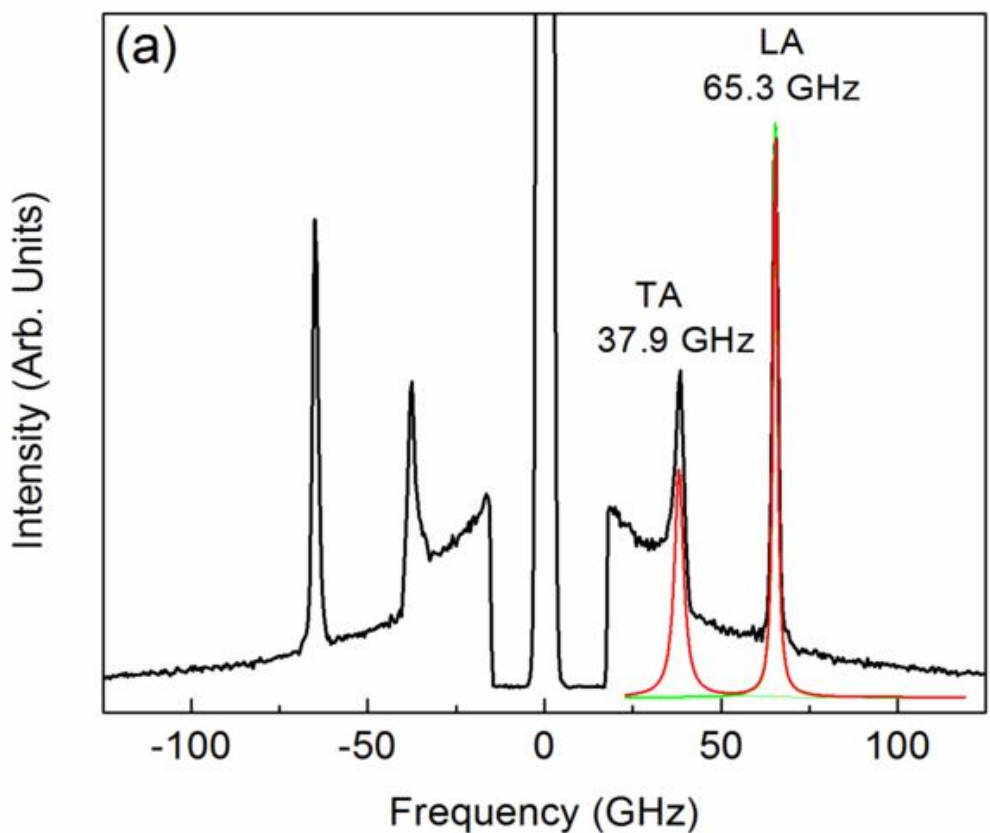
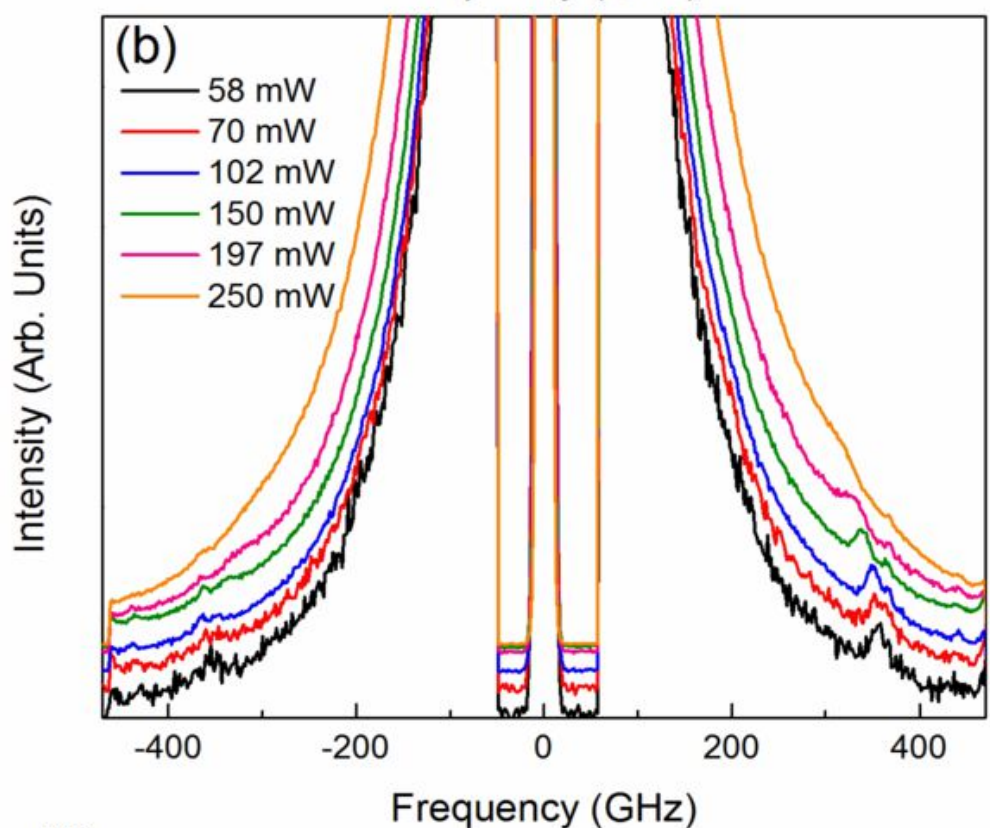
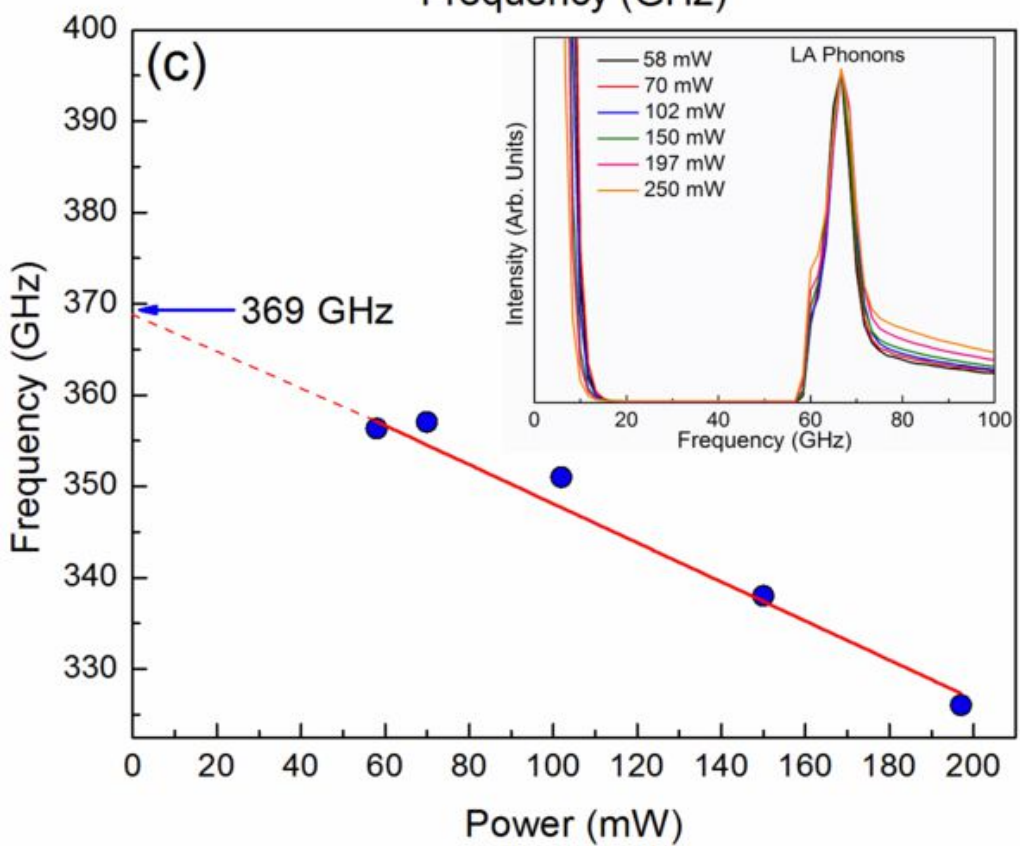

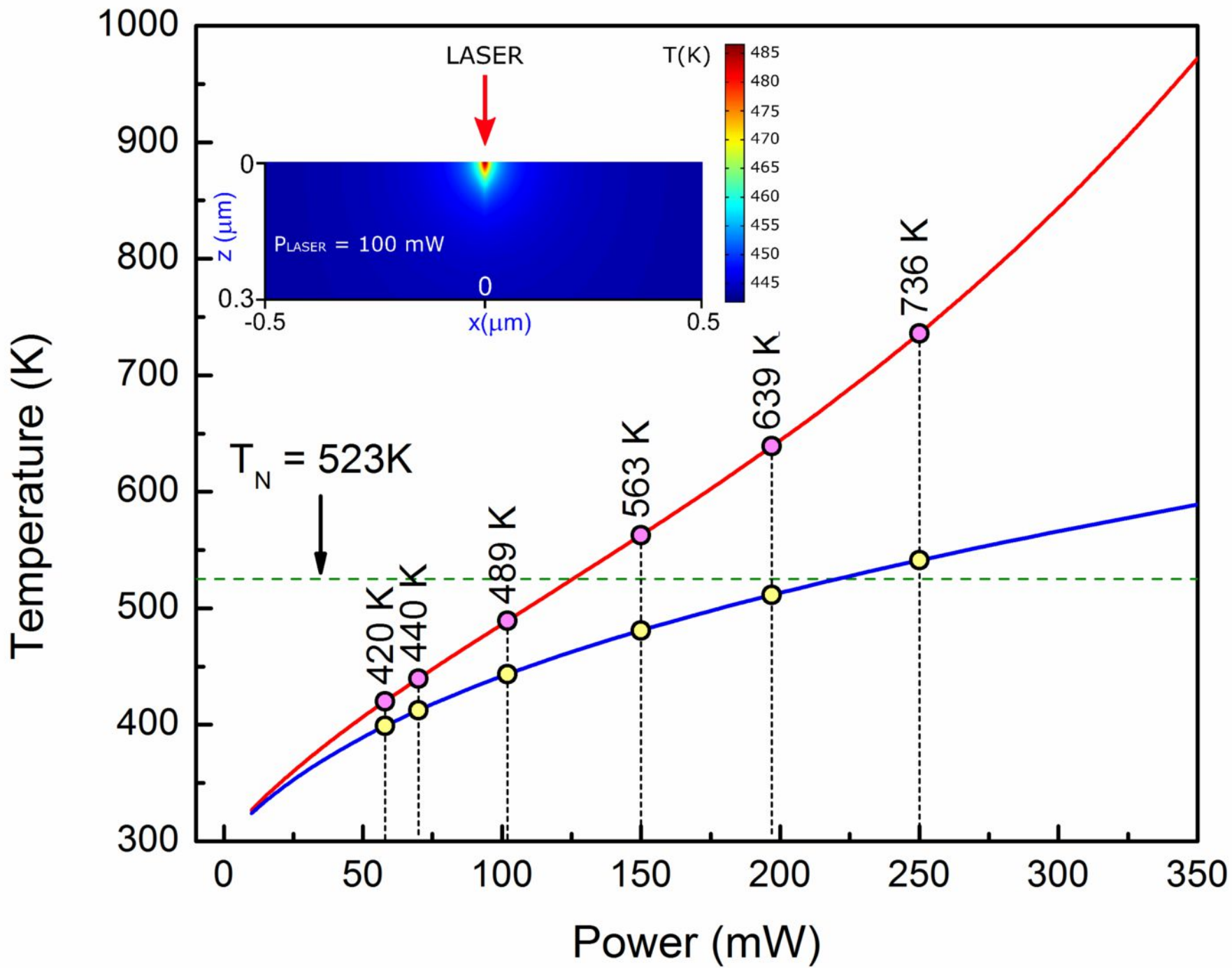